\documentclass[12pt,twoside]{article}
\usepackage{fleqn,espcrc1}

\usepackage{graphicx}
\usepackage[figuresright]{rotating}

\newcommand{\lton}{\mathrel{\lower.9ex
                  \hbox{$\stackrel{\displaystyle <}{\sim}$}}}

\newcommand{\AmS}{{\protect\the\textfont2
  A\kern-.1667em\lower.5ex\hbox{M}\kern-.125emS}}

\title{Microscopic Reaction Dynamics at SPS and RHIC}

\author{Steffen A. Bass\address{Department of Physics,
Duke University, Durham, NC 27708-0305\\
RIKEN-BNL Research Center, Brookhaven National Laboratory,
Upton, NY11973}
}
       
\begin{document}

% typeset front matter
\maketitle

\begin{abstract}
\end{abstract}

\section{Transport Theory at RHIC}

Transport Theory offers the unique capability of connecting experimentally
observable quantities in a relativistic heavy ion collision with its
time evolution and reaction dynamics, thus giving crucial insights
into the possible formation of a transient deconfined phase of
hot and dense matter, the Quark-Gluon-Plasma 
(for reviews on QGP signatures, see e.g.
\cite{Bass:1999vz,harris96a}).

\begin{figure}[t]
%\framebox[172mm]{
\includegraphics[width=160mm]{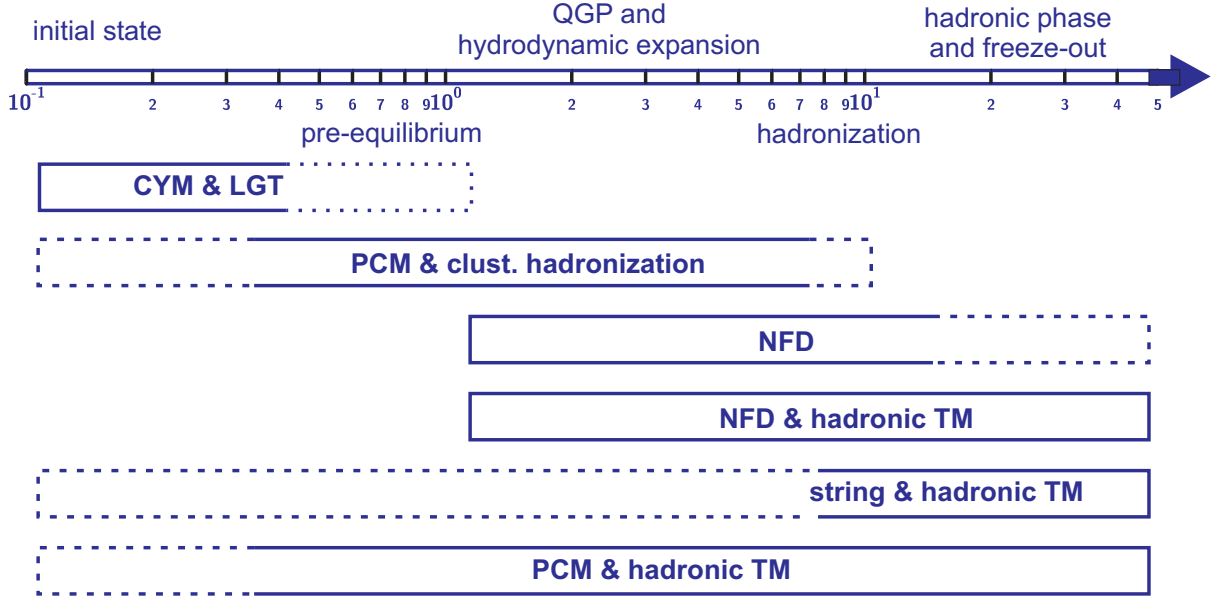}
%}
\vspace{-10mm}
\caption{\label{ttover}
Transport theory approaches for RHIC and their
range of applicability.}
\end{figure}

Figure~\ref{ttover} provides an overview
of different transport theoretical ansatzes currently on the
market for the description of a relativistic heavy ion collision at RHIC
energies. The timeline shows a {\em best case scenario} for what to expect: 
the formation of a QGP with subsequent hadronization and freeze-out.
Bands with solid lines denote the safe range of applicability
for the respective transport approach, whereas dashed/dotted bands 
refer to areas in which the approach is still applied but where the
assumptions on which the approach is based upon may be questionable
or not valid anymore.

The {\em initial state} and early {\em pre-equilibrium  phase} are best described
in Classical Yang-Mills theory (CYM) \cite{MLV}
or Lattice Gauge Transport (LGT) \cite{BMP}
calculations - only these classes of models treat the coherence of the
initial state correctly, but do not provide any meaningful dynamics for
the later reaction stages.

The Parton Cascade Model (PCM) \cite{PCM} and related pQCD approaches
\cite{hijing}
treat the initial state as incoherent parton configuration,
but are very well suited for the {\em pre-equilibrium phase} and subsequent
thermalization, leading to a {\em QGP and hydrodynamic expansion}. 
Microscopic degrees of freedom in this ansatz are quarks and gluons
which are propagated according to a Boltzmann Equation with a collision
term using leading order pQCD cross sections.
Augmented
with a cluster hadronization ansatz the PCM is applicable up
to {\em hadronization}. The range of this model can be even further extended
if it is combined with a hadronic cascade which treats the {\em hadronic phase
and freeze-out} correctly \cite{bass_vni}.

Nuclear Fluid Dynamics (NFD, see e.g. \cite{Bj,DumRi})
is ideally suited for the {\em QGP and hydrodynamic expansion} reaction phase,
but breaks down in the later, dilute, stages of the reaction when the
mean free paths of the hadrons become large and flavor degrees of freedom
are important. The reach of NFD can also be extended by combining it
with a microscopic hadronic cascade model -- this kind of hybrid approach
(dubbed {\em hydro plus micro}) was pioneered in \cite{hu_main} 
and has been now also taken up by other groups \cite{Teaney:2000cw}. 
It is to date the most successful approach
for describing the soft physics at RHIC. The biggest advantage of NFD is
that it directly incorporates an equation of state as input - one of it's
largest disadvantages is that it requires thermalized initial conditions
and one is not able to do an ab-initio calculation.

Last but not least  string and
hadronic transport models \cite{rqmd,urqmd} 
have been very successful in the AGS and SPS energy domains.
They treat the early reaction phase as a superposition of hadron-hadron
strings and are thus ill suited to describe the microscopic reaction dynamics
of deconfined degrees of freedom. In the later reaction stages, however,
they are the best suited approach, 
since they incorporate the full spectrum of degrees of freedom of a hadron
gas (including flavor dependent cross sections).

It is thus important to note that there is not a single transport 
theoretical ansatz currently available, which is able to
cover the entire time-evolution of a collision at RHIC in one self-consistent
approach.
                                                   
\section{Kinetic Evolution}

\begin{figure}[t]
\begin{minipage}[t]{80mm}
\includegraphics[width=85mm]{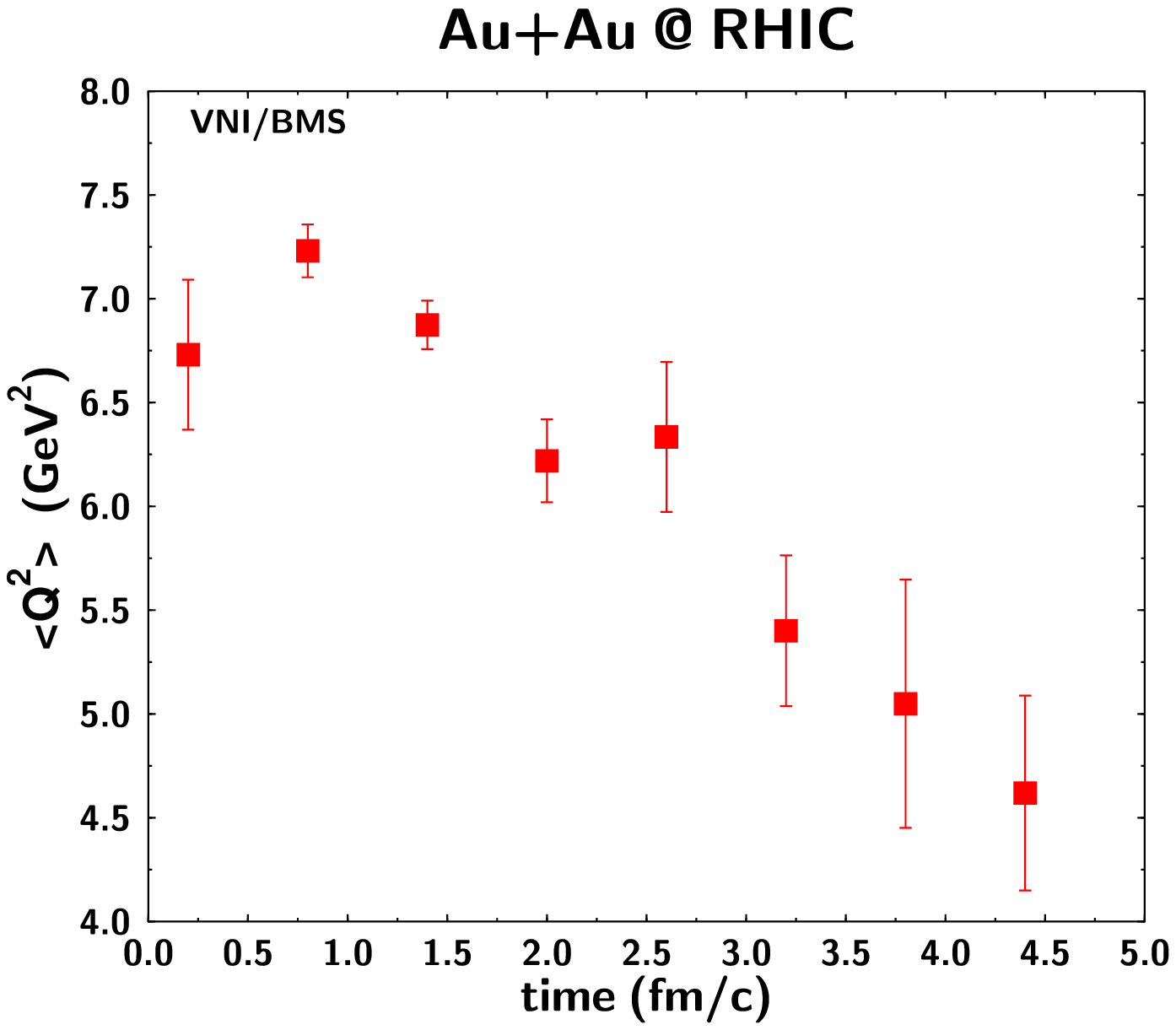}
\vspace{-15mm}
\caption{\label{q2tevol}
Time evolution of parton scattering $\langle Q^2\rangle$ for
central Au+Au collisions at RHIC in the Parton Cascade Model}
\end{minipage}
\hspace{\fill}
\begin{minipage}[t]{75mm}
\includegraphics[width=80mm]{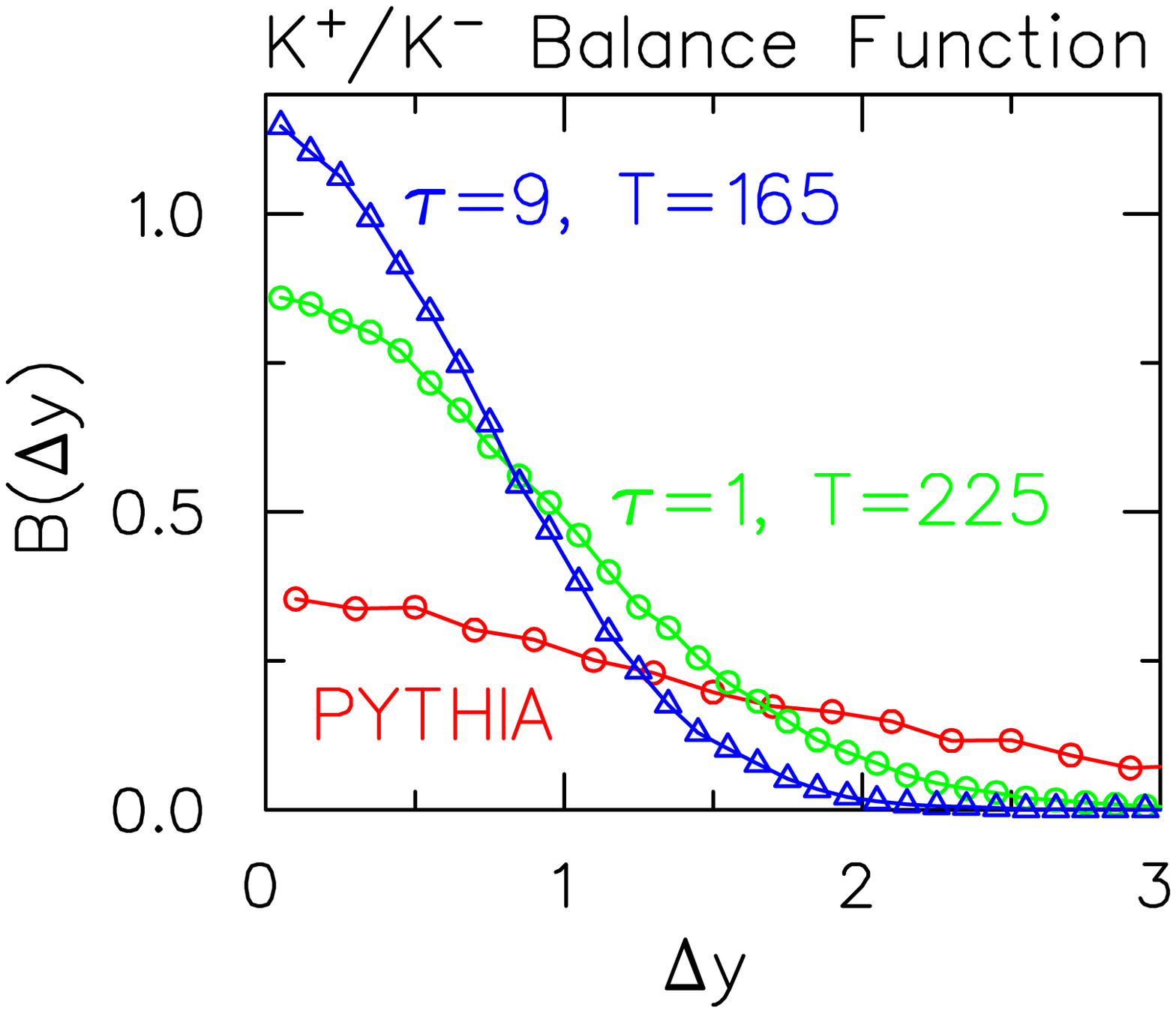}
\vspace{-15mm}
\caption{\label{balancefig}
$K^+/K^-$ Balance Functions for $T_C=165~$MeV, $T_C=225~$MeV and
in a hadronizing string picture}
\end{minipage}
\end{figure}

The initial reaction dynamics are best described by
partonic degrees of freedom.
Figure~\ref{q2tevol} shows the time-evolution of
the average squared momentum transfer $\langle Q^2 \rangle$ 
in parton-parton collisions for central
Au+Au reactions at RHIC, based on a PCM calculation \cite{BMS}.
Early rescatterings
of incident partons occur at rather large values of $Q^2 \approx 7$~GeV$^2$.
However, between the time $t_{\rm c.m.} \approx 0.5$~fm/c and
$t_{\rm c.m.} \approx
4.5$~fm/c the average $\langle Q^2 \rangle$ decreases rapidly 
to a value of about $4.5$~GeV$^2$ until hadronization sets in.
The strong time-dependence of $\langle Q^2 \rangle$ shows that there 
is no single $Q^2$ scale which can be chosen to unambiguously set the 
pQCD scale in an ultra-relativistic heavy-ion collision.
The reaction dynamics in the partonic phase is dominated by
gluon-gluon scattering ($\approx 43$\%) followed gluon fusion
($\approx 27$\%) and quark-gluon scattering ($\approx 26$\%). Quark-quark
and quark-antiquark scattering only contribute on a few \% level. 
These numbers are fairly independent of the system size and change
only marginally between proton-proton and Au+Au reactions.

While there is no doubt about the importance of partonic degrees of
freedom for the initial, early, reaction stages, the crucial question
is, how long the deconfined state actually existed and whether this
time-span is long enough for thermalization and collective effects
to occur. Balance functions offer a unique
model-independent formalism to probe the time-scales of a deconfined phase
and subsequent hadronization \cite{balance}.
Late-stage production of quarks could be attributed to three mechanisms:
formation of hadrons from gluons, conversion of the non-perturbative vacuum
energy into particles, or hadronization of a quark gas at constant
temperature. Hadronization of a quark gas should approximately conserve the net
number of particles due to the constraint of entropy conservation. Since
hadrons are formed of two or more quarks, creation of quark-antiquark pairs
should accompany hadronization. All three mechanisms for late-stage quark
production involve a change in the degrees of freedom. Therefore, any signal
that pinpoints the time where quarks first appear in a collision would provide
valuable insight into understanding whether a novel state of matter has been
formed and persisted for a substantial time.

The link between balance functions and the time at which quarks are created has
a simple physical explanation. Charge-anticharge pairs are created at the same
location in space-time, and are correlated in rapidity due to the strong
collective expansion inherent to a relativistic heavy ion collision. Pairs
created earlier can separate further in rapidity due to the higher initial
temperature and due to the diffusive interactions with other particles. The
balance function, which describes the momentum of the accompanying
antiparticle, quantifies this correlation. 
The balance function describes the conditional
probability that a particle in the bin $p_1$ will be accompanied by a particle
of opposite charge in the bin $p_2$:
\vspace{-3mm}
\begin{equation}
\label{balancedef_eq}
B(p_2|p_1) \equiv \frac{1}{2}\left\{
\rho(b,p_2|a,p_1)-\rho(b,p_2|b,p_1) +\rho(a,p_2|b,p_1)-\rho(a,p_2|a,p_1)
\right\},
\end{equation}
\vspace{-3mm}
where $\rho(b,p_2|a,p_1)$ is the conditional probability of observing a
particle of type $b$ in bin $p_2$ given the existence of a particle of type $a$
in bin $p_1$.  The label $a$ might refer to all negative kaons with $b$
referring to all positive kaons, or $a$ might refer to all hadrons with a
strange quark while $b$ refers to all hadrons with an antistrange quark. 

Figure~\ref{balancefig} shows $K^+/K^-$ balance functions 
as predicted in a simple
Bjorken thermal model for two hadronization temperatures, 
225 MeV and 165 MeV as well as for fragmenting strings (utilizing PYTHIA), 
which would represent the hadronization scenario of a hadronic/string picture
similar to that of RQMD or UrQMD.
Since particles from cooler systems have smaller thermal velocities,
they are more strongly correlated in rapidity and result in narrower balance
functions. A strong sensitivity to the hadronization temperature 
and time can be clearly
observed.

\section{Flavor Evolution}

\begin{figure}[t]
\begin{minipage}[b]{80mm}
\includegraphics[width=75mm]{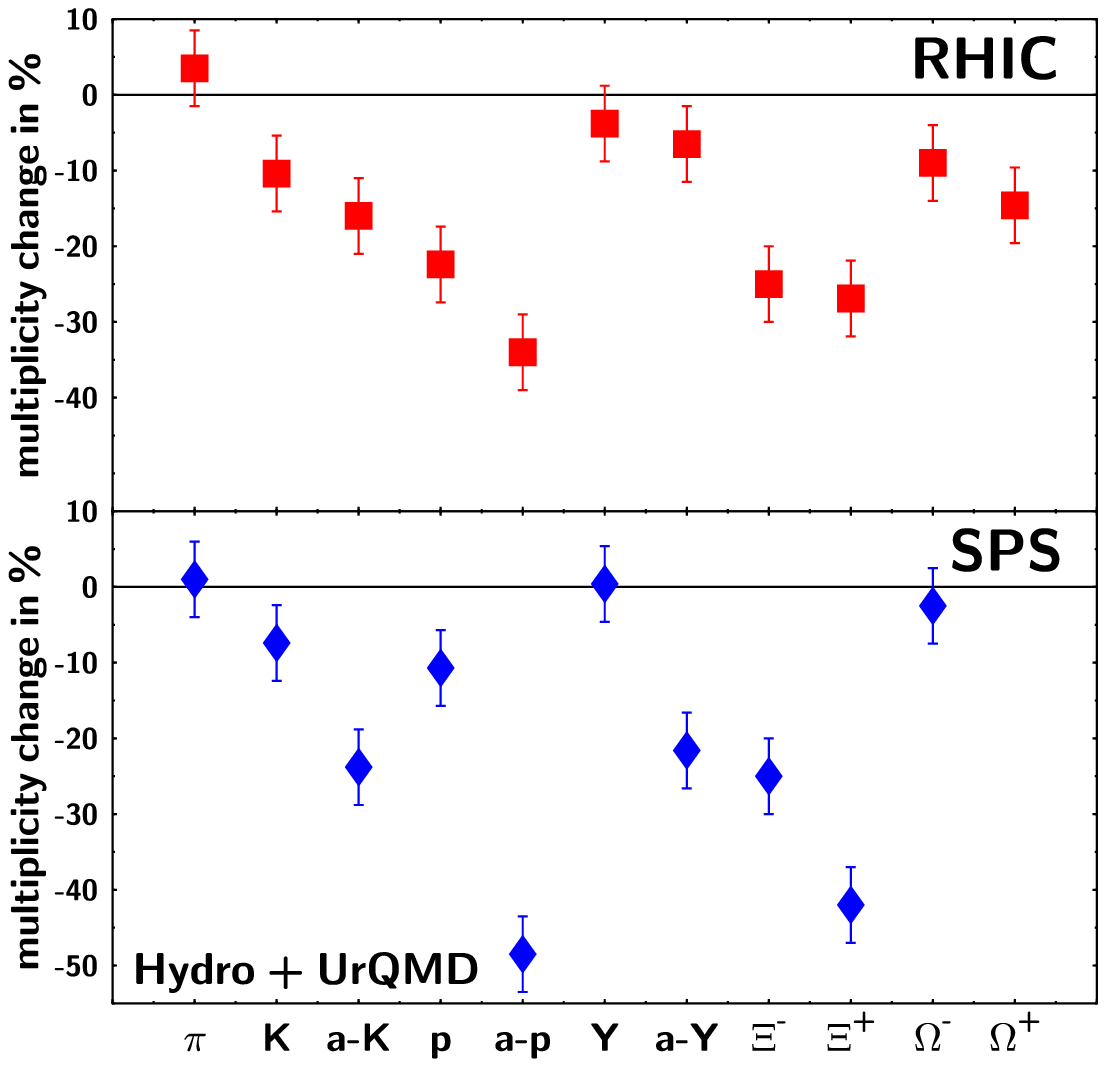}
\vspace{-15mm}
\caption{\label{chem2}
Multiplicity change in \% due to hadronic
rescattering for various hadron species at SPS and RHIC.
The error-bars give an estimate of the systematic error. }
\end{minipage}
\hspace{\fill}
\begin{minipage}[b]{75mm}
\includegraphics[width=80mm]{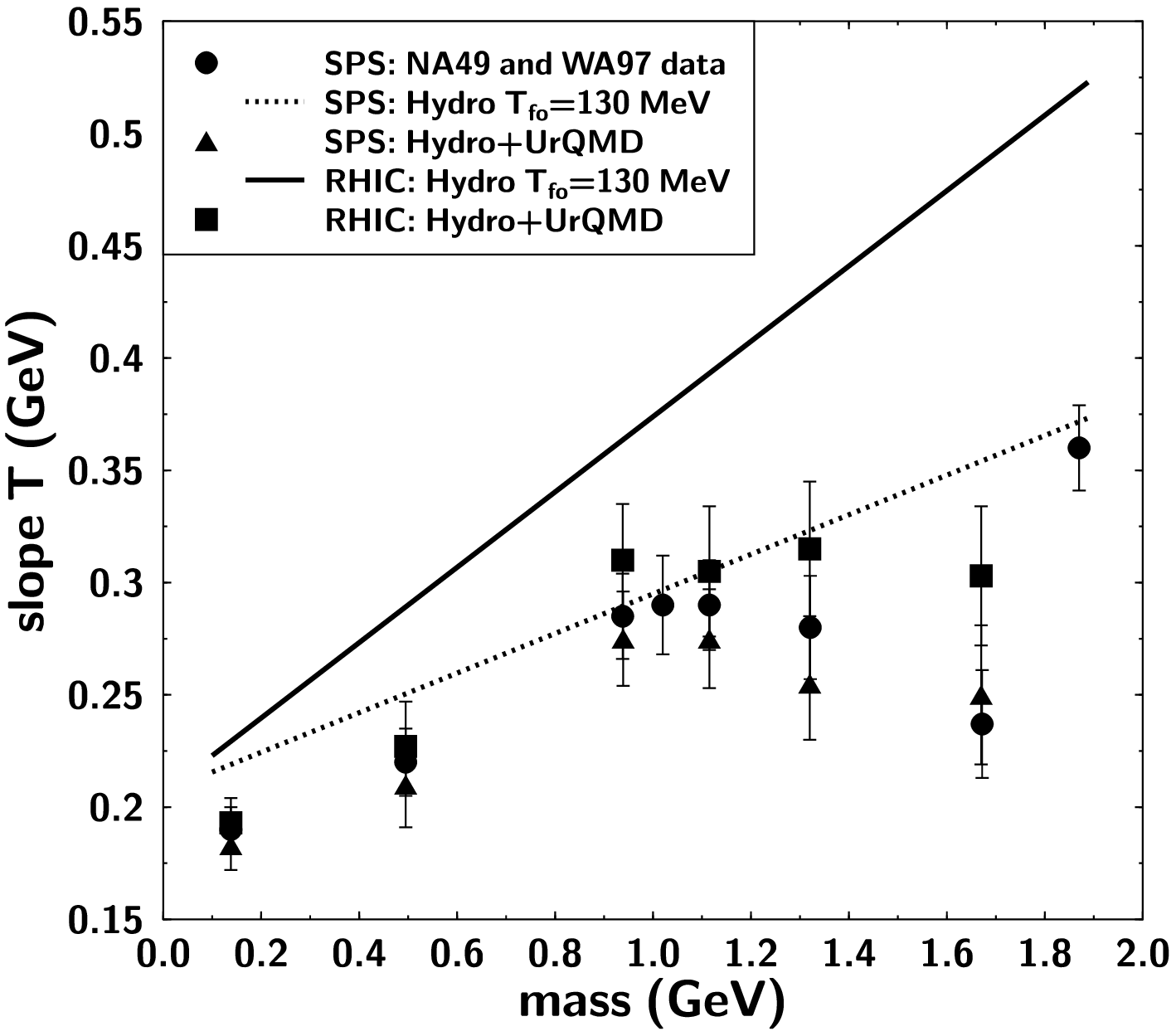}
\vspace{-15mm}
\caption{\label{mslope}
Inverse hadron $m_T$ slopes at $y_{c.m.}=0$.
The lines depict pure hydrodynamics whereas the symbols
refer to data and hydro+micro calculations.}
\end{minipage}
\end{figure}       

A central issue for the characterization of the deconfined phase on
the basis of the measurable final state hadrons is how much the
hadronic composition of the system changes due to hadronic rescattering:
figure~\ref{chem2} shows the relative change (in \%) of the
multiplicity for various hadron species at SPS and RHIC
between hadronization and freeze-out.
As is to be expected, the state of rapid expansion
prevailing at hadronization does not allow chemical equilibrium
to hold down to much lower temperatures. The hadronic rescattering changes
the multiplicities by less than a factor of
two, cf.\ also~\cite{Stock:1999hm}. Thus, we have first evidence
that a QGP expanding and hadronizing as an ideal fluid produces a too
rapidly expanding background for a hadron-fluid with known
elementary cross-sections to maintain chemical equilibrium down to
much lower temperatures than $T_C$.

However, a closer look provides more insight into the
chemical composition. The changes are most pronounced
at the SPS, were the baryon-antibaryon asymmetry is highest (since
the net-baryon density at mid-rapidity is highest). This manifests
e.g.\ in a reduction of the antiproton multiplicity by 40-50\% due
to baryon-antibaryon annihilation. $\bar{\Lambda}$ and $\bar{\Xi}$
are affected in similar fashion.    
It remains a matter of debate whether these changes in the hadronic
yields are consistent with the assumption of chemical freeze-out at
the phase-boundary.

\begin{figure}[t]
\begin{minipage}[t]{75mm}
\includegraphics[width=70mm]{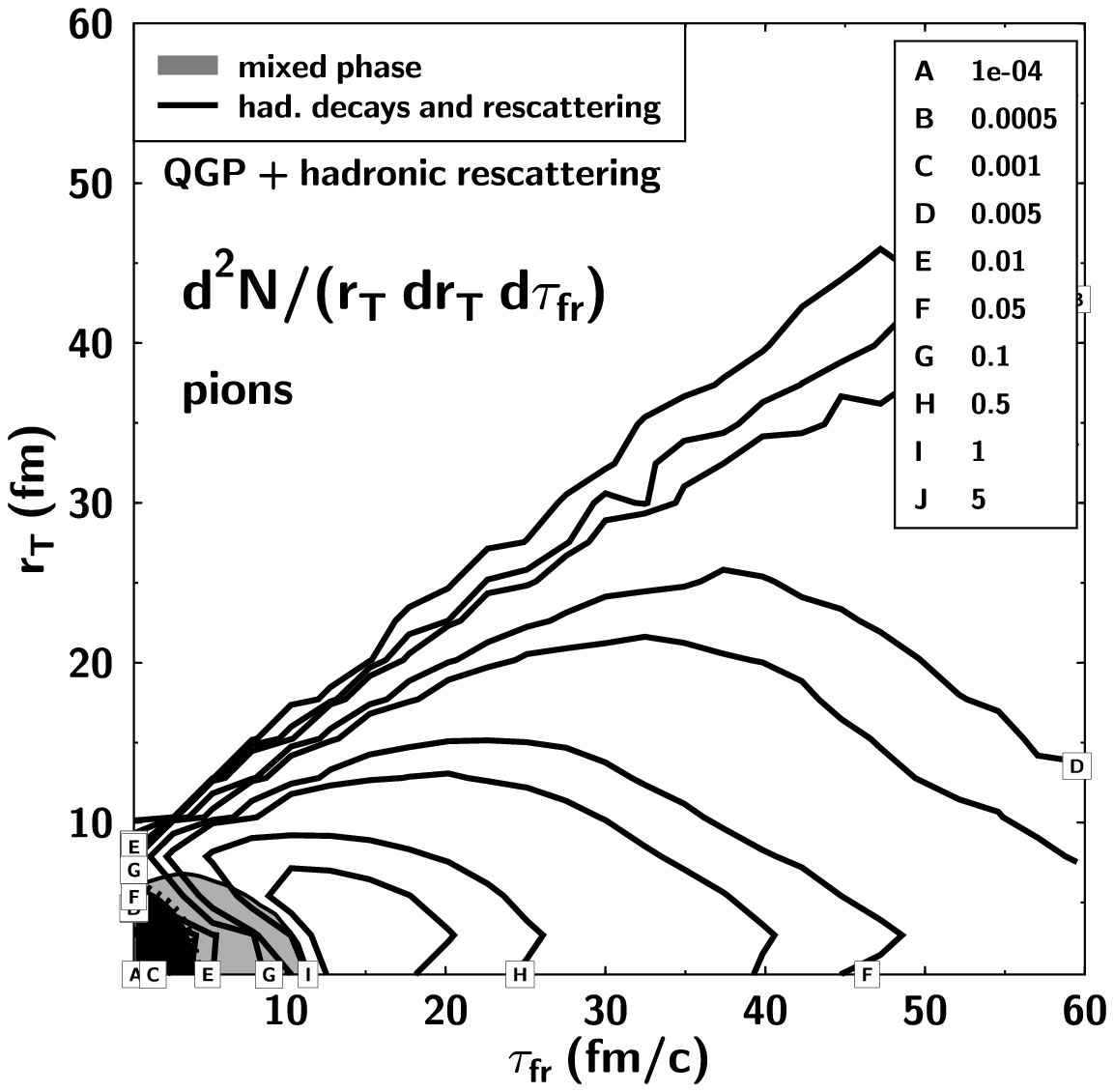}
\vspace{-10mm}
\caption{\label{rttf}
Freeze-out transverse radius and time distribution
$d^2N/(r_Tdr_Tdt_{fr})$ for pions at RHIC.}
\end{minipage}
\hspace{\fill}
\begin{minipage}[t]{80mm}
\includegraphics[width=85mm]{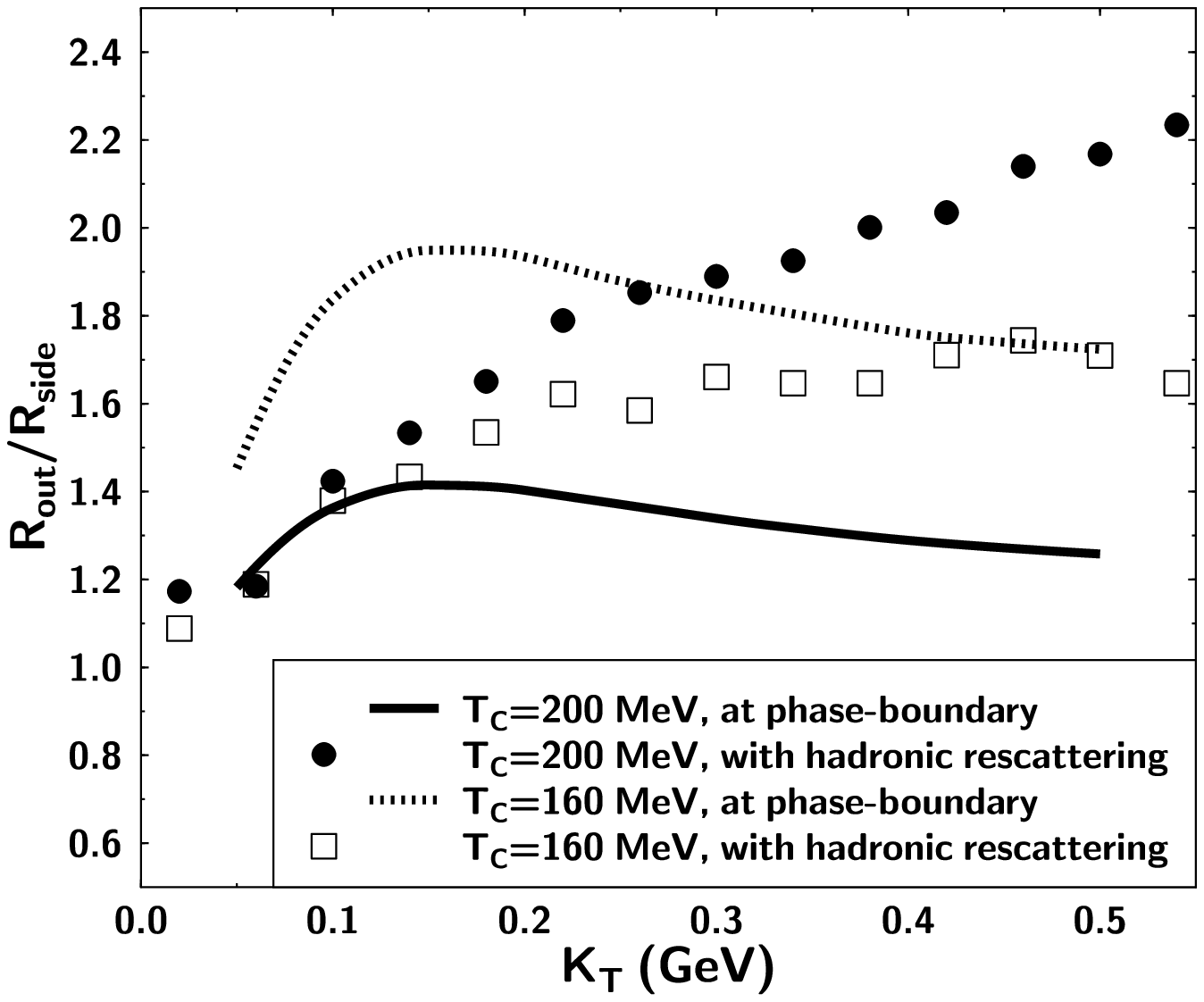}
\vspace{-15mm}
\caption{\label{osratio}
$R_{\rm out}/R_{\rm side}$ for RHIC initial conditions,
as a function of $K_T$ at freeze-out
(symbols) and at hadronization (lines).}
\end{minipage}
\end{figure}

However, despite the large amount of rescattering occuring in the
hadronic phase, some hadron species may yet be sensitive to conditions
at hadronization:
figure~\ref{mslope}
displays the inverse slope parameters $T$ obtained by an
exponential fit to
$dN_i/d^2m_Tdy$ in the range $m_T-m_i<1$~GeV for SPS and RHIC in a hybrid
hydro+micro calculation and
compares them to SPS data \cite{strange_data}.
The trend of the data, namely the ``softer'' spectra of $\Xi$'s and
$\Omega$'s as compared to a linear $T(m)$ relation is reproduced
reasonably well. This is in contrast to ``pure'' hydrodynamics with kinetic
freeze-out on a common hypersurface (e.g.\ the $T=130$~MeV isotherm), where
the stiffness of the spectra increases linearly with mass
as denoted by the lines in
fig.~\ref{mslope}.
When going from SPS to RHIC energy, the model discussed here generally
yields only a slight increase of the inverse slopes,
although the specific entropy is larger by a factor of 4-5~!
The reason for this behavior is the first-order phase transition that
softens the transverse expansion considerably.   

The softening of the spectra is caused by
the hadron gas emerging from the hadronization of the QGP
being almost ``transparent'' for the multiple
strange baryons.
This can be seen by calculating the average number of collisions
different hadron species suffer in the hadronic phase:
whereas $\Omega$'s suffer on average only one hadronic interaction,
$N$'s and $\Lambda$'s suffer approx. 5--6
collisions with other hadrons before they freeze-out.

Thus, one may conclude that
the spectra of $\Xi$'s and especially $\Omega$'s  are practically
unaffected by the hadronic reaction stage and closely resemble
those on the phase
boundary. They therefore act as probes of the QGP expansion prior
to hadronization and can be used to measure the expansion rate
of the deconfined phase.               

\section{Freeze-out}

Kinetic freeze-out, at which all (elastic) interactions cease, terminates 
the evolution of the reaction discussed in figure~\ref{ttover}. However,
the freeze-out time of the system is ill-defined, since interactions
cease locally on a particle per particle basis.
Fig.~\ref{rttf} shows the distribution of freeze-out points of pions
in the forward light-cone, as well
as the hadronization hypersurface from where all hadrons emerge in a 
hydro+micro approach.
Freeze-out is seen to occur in a {\em four-dimensional} region within the
forward light-cone rather than on a three-dimensional ``hypersurface'' in
space-time. Similar results have
also been obtained within other microscopic transport models~\cite{microFO}
when the initial state was not a QGP.
It is clear that the hadronic system disintegrates  slowly (as
compared to e.g.\ the hadronization time), rather than emitting a ``flash''
of hadrons (predominantly pions) in an instantaneous decay.

A first order phase transition
leads to a prolonged hadronization time as compared to a cross-over
or ideal hadron gas with no phase transition, and has been related to unusually
large Hanbury-Brown--Twiss (HBT) radii~\cite{pratt86,schlei,dirk1}.
The phase of coexisting hadrons and QGP reduces the ``explosivity'' of the
high-density matter before hadronization, extending the emission duration of
pions~\cite{pratt86,schlei,dirk1}.
This phenomenon should then depend on the
hadronization (critical) temperature $T_c$ and the latent heat of the
transition. For recent reviews on this topic we refer
to~\cite{reviews,wiedemannrep}.

It has been suggested that the ratio $R_{\rm out}/
R_{\rm side}$ should increase strongly once the initial
entropy density $s_i$ becomes substantially larger than that of the hadronic
gas at $T_c$~\cite{dirk1}.
The strong $T_C$ dependence
of $R_{\rm out}/R_{\rm side}$ in such a purely hydrodynamical scenario
can be seen in fig.~\ref{osratio} (solid and dotted lines). 
Here, the $R_{\rm out}/R_{\rm side}$ ratio at hadronization (calculated
in a hydrodynamical scenario) is compared 
to the ratio after subsequent hadronic rescattering and freeze-out 
in a combined hydro+micro approach \cite{hbt_prl}. 
Clearly, up to $K_T\sim200\,$MeV
$R_{\rm out}/R_{\rm side}$ is independent of $T_c$, 
if hadronic rescatterings are taken into account.
Moreover, at higher $K_T$ the dependence on $T_c$ is even reversed:
for high $T_c$ the $R_{\rm out}/R_{\rm side}$ ratio even exceeds that
for low $T_c$.
Higher $T_c$ speeds up hadronization but on the other hand prolongs
the dissipative hadronic phase that dominates the HBT radii. This is because
during the non-ideal hadronic expansion the scale of spatial homogeneity
of the pion density distribution increases, as the pions fly away from the
center, but the transverse flow can hardly increase to counteract.
Therefore, after hadronic rescattering
$R_{\rm out}/R_{\rm side}$ does not drop towards
higher $K_T$ (in the range $K_T\lton 3m_\pi$).

For central collisions of Au nuclei at
$\sqrt{s}=130A$~GeV, preliminary data from STAR gives
$R_{\rm out}/R_{\rm side}\simeq1.1$ at small $K_T$~\cite{Lisa00}.
Results from RHIC at higher $K_T$ will test whether a long-lived
{\em hadronic} soft-rescattering stage, associated with the formation and
hadronization of an equilibrated QGP state is indeed seen in
heavy ion collisions at the highest presently attainable energies. 

I wish to thank my collaborators A. Dumitru, P. Danielewicz, 
B. M\"uller, S. Pratt, S. Soff, D. Srivastava and
the UrQMD collaboration, who have 
significantly contributed to the results reviewed in this article.
This work was supported by DOE grant DE-FG02-96ER40945.

\end{document}